# Hardware in Loop Learning with Spin Stochastic Neurons

*A N M Nafiul Islam, Kezhou Yang, Amit Kumar Shukla, Pravin Khanal, Bowei Zhou, Wei-Gang Wang, Abhronil Sengupta\**

A. N. M. N. Islam, A. K. Shukla[†], A. Sengupta

School of Electrical Engineering and Computer Science, The Pennsylvania State University, University Park, PA 16802, USA

E-mail: sengupta@psu.edu

[†]Current Affiliation: GlobalFoundries, India.

K. Yang[†], A. Sengupta

Department of Materials Science and Engineering, The Pennsylvania State University, University Park, PA 16802, USA

[†]Current Affiliation: Hong Kong University of Science and Technology, Guangzhou, China.

P. Khanal, B. Zhou, W. G. Wang

Department of Physics, University of Arizona, Tucson, AZ 85721, USA

Keywords: hardware-in-loop, spintronics, stochastic neuron, neuromorphic devices, edge-intelligence

**Abstract**

Despite the promise of superior efficiency and scalability, real-world deployment of emerging nanoelectronic platforms for brain-inspired computing have been limited thus far, primarily because of inter-device variations and intrinsic non-idealities. In this work, we demonstrate mitigating these issues by performing learning directly on practical devices through a hardware-in-loop approach, utilizing stochastic neurons based on heavy metal/ferromagnetic spin-orbit torque heterostructures. We characterize the probabilistic switching and device-to-device variability of our fabricated devices of various sizes to showcase the effect of device dimension on the neuronal dynamics and its consequent impact on network-level performance. The efficacy of the hardware-in-loop scheme is illustrated in a deep learning scenario achieving equivalent software performance. This work paves the way for future large-scale implementations of neuromorphic hardware and realization of truly autonomous edge-intelligent devices.

# 1. Introduction

Interest in bio-plausible devices and systems stems from the brain's unique ability to process real-world information effectively and efficiently. Although deep artificial neural networks have recently been able to come close and even surpass human-level performance in some cases, the energy and area cost associated are still many-folds over their biological counterparts.[1] Thus, in-memory computing paradigms akin to the brain designed with specialized electronics for intrinsic emulation of neuronal and synaptic functionalities have emerged as alternatives to the traditional von-Neumann architecture and complementary metal oxide semiconductor (CMOS) technologies.[2–7] However, their adoption in large-scale neuromorphic hardware is scarce[8], relying rather on CMOS.[9,10] A key factor behind this has been device-to-device variations and inherent non-idealities of these emerging devices[7,11–13], requiring additional peripheral circuitry for reliable operation, eroding their area and energy advantage.

While there have been efforts in literature to characterize device non-idealities and their impact on learning for artificial synapses of varied technologies[7,14,15], such studies for artificial neurons, specifically stochastic ones, have been rather limited.[16] The robust computational ability of the brain is largely attributed to its noisy probabilistic nature.[17] Additionally, to reach brain-like memory densities, continual scaling of these devices is needed. Although the energy and area advantages of scaling in CMOS is evident, it is not so straightforward for neuromorphic devices; non-idealities can become an insurmountable issue. Thus, for large-scale integration, it is pivotal to understand the interplay between device size and its impact on the variation and stochasticity of the neuronal dynamics. A systematic analysis through characterization of realistic devices identifying their categorical effect on network-level performance for deep learning applications has been missing. Similarly, various schemes for variation compensation exist, however, their potency has not been tested in experimental demonstrations for deep learning tasks such as pattern recognition.

In this work, we demonstrate such an investigation for spintronic stochastic neurons based on a heavy metal/ferromagnetic spin-orbit torque (SOT) hall-bar heterostructures[18] along with a proof-of-concept demonstration of hardware-in-loop learning offsetting intrinsic hardware variations for deep learning applications. We study the sigmoidal stochastic switching of the SOT devices and its variances for a wide range of device sizes, ranging from 5µm to 300nm. We find that the necessary bias current for switching decreases with decreasing size while the slope of the probabilistic switching characteristics increases, highlighting an intertwined inverse relationship between power consumption, accuracy, and robustness in neural network scenarios. Finally, we extrapolate the results of our experimental hardware-in-loop setup to draw insights for large-scale neuromorphic implementations through a hardware-software co-analysis.

Spintronic devices with their nanosecond response capabilities, and compatibility with existing nanoelectronics are a great prospect for realizing neuromorphic frameworks.[19] Compared to other emerging technologies such as phase change, and resistive memories[7,20], spintronic devices are more compact and require less operating energy[21–24] – thereby resulting in substantial energy savings at the system level.[25] Prior works looking at the device-to-device variations of spin-based devices have either looked at them for inference only[26] or have not exploited the device stochasticity.[27] Additionally, previous stochastic hardware implementations and hardware-in-loop learning have focused more on probabilistic computing applications[28,29] and associative learning.[27,30] The more powerful and widely used neural networks employing backpropagation for training have only been demonstrated in software.[16,31] Moreover, these demonstrations have all lacked comprehensive characterization analysis of device properties. Our work tries to overcome all these issues through extensive experimental characterization of devices of varied dimensions and hardware demonstration of a cognitive task, namely recognition of handwritten digits, to present the true potential of stochastic spintronic devices.

## 2. Results

### 2.1. Realizing spin-orbit torque based stochastic neurons

In biological neural networks, neurons serve as the key computing unit. As shown in **Figure 1**(a), inputs are propagated from pre-synaptic neurons to post-synaptic neurons, which integrate them and fire output spikes once a certain threshold is crossed. Neurons, particularly in the cortex, have been observed exhibiting stochastic firing with nonlinear dependence on the resultant post-synaptic current input to the neuron.[32] In this work, we realize such stochastic neurons displaying non-linear dependence to input current with spintronic devices employing spin-orbit torque. The core device structure, shown in Figure 1(b), consists of a heavy metal (HM) layer and a ferromagnetic (FM) layer. When a charge current flows through the HM/FM structure, spin current is generated due to primarily two effects – 1) Spin Hall effect in the HM and 2) the Rashba effect at the FM/HM interface.[33,34] The spin Hall effect originates from both intrinsic sources such as band structures and extrinsic sources like Mott scattering by impurities.[35,36] Additionally, electrons moving in the interfacial electric field at HM/FM interface experience a magnetic field, known as the Rashba field, which introduces a spin-orbit term in the system Hamiltonian. The induced spin-orbit interaction splits the band of different spins, which enables non-zero net spin accumulation under applied electric field.[37–39] The generated spin current induces FM magnetization switching when charge current reaches switching current.[40] Due to thermal noise, the switching is probabilistic rather than deterministic when pulse current is applied. Thus, at any given time, the switching probability depends on the magnitude of the pulse current, which follows a sigmoidal function. This provides the non-linearity required for the operation of the neuron. We, thus, model the activation of neurons, $a$, as follows:

$$a = \begin{cases} 1 \text{ with probability}, p_1 = sigmoid(z) = \dfrac{1}{1 + e^{-z}} \\ 0 \text{ with probability}, p_0 = 1 - p_1 \end{cases} \quad (1)$$

Here, $z$ is the input coming to the neuron. Previously, it has been shown that the average firing activity of such a stochastic neuron over time linearly approximates the sigmoid function used for computation in traditional neural networks.[16] Going beyond inference,[41] featured such a stochastic sigmoid used directly in training neural networks to eradicate the need for evaluating the network over time.

The fabricated device and material stack is shown in Figure 1(c). The details of the fabrication process and device geometry can be found in the Methods Section and Supporting Information (SI) Figure S2, respectively. A Hall bar structure is used, so that the device magnetization can be probed out using the anomalous Hall effect, where a voltage difference occurs across terminals perpendicular to the flow of current by accumulation of electrons with different spin directions.[36] A magnetic field in out-of-plane direction is applied to obtain the hysteresis loop shown in Figure 1(d). The rectangle hysteresis loop indicates perpendicular magnetic anisotropy (PMA).

### 2.2. Effect of dimension on device characteristics

In order to characterize the switching and quantify the effect of dimension reduction, experiments are conducted on devices with bar widths of 5.0µm, 2.5µm, 2.0µm, 1.5µm, 1.0µm, 0.9µm, 0.7µm, 0.5µm and 0.3µm. For each size, we studied the switching behavior of 4 different devices to identify the device-to-device variability. The measurement setup is shown in **Figure 2**(a). A 200 Oe in-plane magnetic field along the direction of current flow (x-direction) is applied. We begin by confirming the SOT induced magnetization switching of the devices by applying gradually increasing pulses. Again, the state of the device is probed by measuring the anomalous Hall resistance, $R_{AHE}$, shown in Figure 2(b). We observe that as the device width is reduced, the hysteresis loop shrinks accordingly, i.e., the devices switch at lower magnitudes of pulse currents. Note, the current switching behavior is stochastic and thus next we characterize this behavior. For this, devices are applied with 100 iterations of reset-set cycles. In each iteration, the devices are first applied with a reset pulse with a pulse width of 100µs, to initialize the device state. The reset pulse must be large so that the device is in the '-1' state. This is confirmed by reading out the $R_{AHE}$. Afterwards, a set or write current pulse, $I_{write}$, with pulse width 100µs is applied to switch the device. The magnitude of $I_{write}$ is increased to figure out the switching dynamics of the devices. We confirm the switching probability of the device shows the expected sigmoidal relationship with pulse amplitude (SI Figure S3(a)). The process of obtaining this relation is detailed in SI Figure S3(b). In Figure 2(c-k), the switching dynamics of 4 individual devices of each of the 9 widths considered is shown. In order to fit the dynamics to that of an ideal sigmoid, the neuronal switching due to current can be thought of having two components:

$$I_{write} = I_{bias} + I_{syn} \qquad (2)$$

Here, $I_{bias}$ is the necessary current to the HM layer of the spin neuron to bias it at 50% probability, whereas $I_{syn}$ is the resultant input synaptic current to the neuron. Note, $I_{syn}$ needs to be normalized by a factor $I_0$, which encodes the degree of dispersion of the neurons' sigmoidal characteristics. Generally, we find that smaller width corresponds to smaller dispersion or programming window. Additionally, we find that the dynamics of each individual spin neuron is quite stable over time (SI Figure S4). We also observe great consistency in performance and endurance (SI Figure S5).

### 2.3. Network evaluation of device performance

Understanding the algorithmic impact of hardware parameters not only allows optimization of performance, power consumption, and reliability, but also enables identification of key design trade-offs. Known as hardware-software codesign, we analyze the interaction between network-level performance metrics and the device properties for our spin stochastic neurons. **Figure 3**(a, b) summarizes the characterization results. We observe that both the switching bias current, $I_{bias}$ and the programming window, which is defined by the pulse amplitude range between 0.01% and 99.9% switching probability, increase linearly with bar width. We found no significant trend between the percentage variation observed and the device bar width. Note, the programming window measures the dispersion of the sigmoid and thus is a function of the normalizing factor, $I_0$ (SI Figure S6). We confirm the linearity of the relationship for bias current and programming window with device width using micromagnetic simulations as well (SI text and SI Figure S7).

We design a hardware-software co-analysis framework to perform a network-level evaluation for our spin stochastic neurons. We perform handwritten digit recognition on the MNIST dataset [42], with a network architecture of 784 input neurons, 400 hidden neurons and 10 output neurons. We chose this prototypical network as it employs backpropagation-based stochastic gradient descent [43] as its learning algorithm, the current standard for deep learning applications. Note, the discussions hold true for any neural network topology. Here, we are only interested in neuronal dynamics and its impact on performance. The network after training achieves an inference accuracy of 97.13% with the spin stochastic neuron dynamics. Details of the network simulation can be found in the Methods Section.

From characterizing the devices, we observe that the device-to-device variation of the bias current required for the devices can be up to 25%. We evaluate the impact of this variation at the system level for the different device dimensions. For the network simulation, the neurons are modelled using the sigmoidal device dynamics. The required bias current for each device is sampled using the following equation:

$$I_{bias}^i = \bar{I}_{bias}(1 + \eta * x) \qquad (3)$$

Here, $\bar{I}_{bias}$ is the average bias current obtained from experimental results, $\eta$ is the percentage of input bias variation and $x$ is a random variable sampled from a uniform distribution between [-1,1] to add the noise or variability. We chose the uniform distribution instead of the standard gaussian or normal distribution, as that represents a worst-case scenario for the network in terms of variability. In Figure 3(c), the result of the analysis is shown. We see that networks simulated with larger devices, i.e., devices with higher average bias current ($\bar{I}_{bias}$), are less prone to this bias variation in contrast to the same network simulated using smaller ones. This is because with the decrease in size, the dispersion of the probability characteristics or the width of the operating current window decreases. Thus, variations force the neuron into its saturation region, depleting its computation abilities. Larger devices with their larger operating windows are less susceptible. This also highlights the intrinsic tolerance abilities of neuromorphic architectures. It should be noted here that we are not considering discretization effects in the synaptic current as spintronic neurons can be directly interfaced with the synaptic crossbars, resulting in reduced peripheral overhead involving Analog-to-Digital converters [44]. Additionally, we find that the number of time-steps required for the spiking network to converge to its final accuracy increases as the percentage of variation increases, thus indicating a trade-off between latency and variation as well (SI Figure S9).

In addition to evaluating variation effects at the system level, another key factor of consideration is interplay of device sizing with system level power consumption. For implementing a full-scale system, the proposed neurons will be interfaced with cross-arrays of synaptic devices where the network weights are mapped to the various conductance levels of the synaptic devices (SI Figure S10). The input currents to our spin neurons are modulated by the conductance. For the weight in-between the $i$-th input and the $j$-th neuron, $w_{i,j}$, the corresponding conductance is given by,

$$G_{i,j} = w_{i,j} G_0 \qquad (4)$$

Here, $G_0$ is the mapped conductance for when the weight is '1'. The synapses can be implemented by any memristive technology, including phase change devices [45,46], spintronic devices [19], ferroelectric devices [6,47], among others. Note, the focus of this article is on neurons and compliments works on these synaptic devices. Our experimental setup consisted of software running on an external computer to mimic the functionality of a synaptic array that was then fed to the set of fabricated stochastic neurons.

To evaluate the impact of neurons and its dimension on such cross-arrays, let us consider the net synaptic current flowing through the HM layer of the neuron. The total current to the $j$-th neuron coming from the synaptic array is given by,

$$I_{syn} = \sum_i G_{i,j} V_i \qquad (5)$$

Here, $V_i$ is the voltage coming in from the $i$-th input. The two factors in $I_{syn}$ are scaled by the mapped conductance for unity weight, $G_0$ and the input spike magnitude, $V_0$ respectively. As the synaptic current has to be scaled to fit the ideal sigmoid by a factor $I_0$, the two has to be equal, giving us, $I_0 = G_0 V_0$. This indicates that the currents to the neuronal devices need to be adjusted in order to ensure proper operation within the programming window. As $I_0$ decreases with width, keeping $G_0$ fixed, i.e., while keeping the conductance mapping the same, the input spike magnitude $V_0$ can be reduced. Decreasing $V_0$ has a major impact on the total power consumption of the network since the number of synapses in a neural network is generally two or three orders of magnitude larger than the number of neurons. Figure 3(d) shows this impact in terms of normalized synaptic read energy required to achieve the target accuracy over 50 time-steps. Scaling down from 5µm to 0.3µm, results in a 50× reduction in energy consumption. Thus, we find that although larger bar widths result in greater tolerance to variation, it also incurs a large energy penalty when looked at from a system integration point of view. Thus, the interplay between the trade-offs needs to be carefully considered when designing such neuromorphic systems, further showcasing the pivotal role of hardware-software codesign.

**2.4. Proof-of-concept Hardware-in-loop Training**

The conspicuous impact of device-to-device variation was highlighted in the previous paragraph with smaller devices being affected more than their larger counterparts. However, scaling is a highly desirable feature as it lowers the total energy cost considerably. Thus, it is essential to overcome these issues. Numerous approaches for repressing these issues have been outlined on the device-level[48–50] and network-level.[51–53] Here, we present proof-of-concept demonstration of a network-level approach where the neuronal devices are included in the training process of a handwritten digit recognition problem, allowing the network to learn the desired patterns with the effect of device-to-device variations included. We perform this through a hardware-in-loop scheme, shown in **Figure 4**(a), where the incoming training images are converted to temporal spikes and are fed into the neuron hardware after modulation through a feedforward network. Based on the input, the spin neurons switch, which is used to calculate the activations and the subsequent errors, gradients and weight update in software and update the synaptic weights of the network. Further details can be found in the Methods Section. It is should be noted that, while such an in-hardware learning scheme does incur high endurance cost, spin-based devices have already been reported in literature with great endurance, over $10^{15}$ cycles.[21]

For the hardware-in-loop learning, we use four spin neuron devices of size 0.5µm. We chose the 0.5µm devices as they would be great candidates that allow for low power operation and also to display the efficacy of the variation compensation scheme. The neuronal dynamics of the four devices are shown in Figure 4(b), along with the fitted sigmoid. As four neurons are used as the output, we train the network on four classes from the MNIST dataset ("0", "2", "4" and "6"), with 4 images from each group. The network architecture is given in SI Figure S11. For testing, we use a single sample from each class. We track the network's training loss during the training process, as illustrated in Figure 4(c). We see that the loss gradually decreases. After training, we use the network for inference on the 4 test images and observe the network input for the 4 hardware neurons. As can be seen from Figure 4(d), the network is able to differentiate between the classes and correctly identify 3 out of the 4 images. The failure to recognize the class "2" image can be attributed to the small size of training set and the apparent dissimilarity between the training samples and test sample of class "2" (SI Figure S11). To compare the hardware-in-loop performance with performance without hardware-in-loop training, we trained the same network in software with the same learning parameters and then used the spin stochastic neurons for inference only. In this scenario, we found that the network can identify only 1 sample correctly (Figure 4(e)). This showcases the efficacy of the hardware-in-loop learning scheme. Additionally, as such neuromorphic hardware systems are expected to be employed in resource-constrained environments, we perform the same experiment with a single neuronal device, time-multiplexed to serve the function of two neurons. Although in such a scenario there will be no concern for device-to-device variations, the hardware non-idealities and cycle-to-cycle variation urges us to confirm the efficacy of such an approach. Here, we again see that such a network can achieve ideal accuracies, further corroborating the need for including hardware in training for edge intelligence applications (SI Figure S12).

## 3. Conclusion

To summarize, we presented a detailed analysis of SOT-based spintronic stochastic neurons and underscored the importance of hardware-software co-design for neuromorphic hardware stressing the interplay between power, accuracy, and robustness for 36 devices of 9 varying sizes, ranging from 5µm to 0.3µm. In total, we corroborated our co-design analysis by conducting close to 100,000 different measurement steps. We demonstrated how these variations can be compensated by in situ learning through a hardware-in-loop learning scheme – showing the potential of variation-prone scaled hardware for edge intelligence applications.

## 4. Methods

### 4.1. Material Deposition and Characterization

The devices in the study are fabricated from thin films deposited in an UHV sputtering deposition with a base pressure in the range of $10^{-9}$ Torr. The stack structure of the films is Si/SiO$_2$(300nm)/Ta (5nm)/CoFeB (1nm)/MgO (2.5nm)/Ta (2.5nm/5nm). The first Ta layer (5nm) serves as the HM layer to provide SOT to the CoFeB (1nm) with PMA. The second Ta layer (2.5nm/5nm) is serving as the capping layer to prevent oxidation of the stack underneath. Detailed fabrication parameters of the thin films can be found in previous publications. [54–58] Hall bars with ≥1µm width have a top capping Ta thickness of 5nm, while bar sizes smaller than 1µm have a top capping Ta thickness of 2.5nm.

### 4.2. Hall-bar Fabrication

The spin stochastic neurons are created using Hall bar structures with the deposited material stack. Bars with ≥1µm width are defined by photolithography, using SPR-3012 as photoresist. Hall bars with bar sizes smaller than 1µm are defined by E-beam lithography, using polymethyl methacrylate (PMMA) as resist layer. The defined patterns are etched by Ar until the Si/SiO$_2$ substrate. The residual resist after etching is removed by soaking in PRS-3000 under 80°C water-bath. Ti (5µm)/Au (100µm) stack are deposited as contact layer by E-beam evaporation after bars are fabricated.

### 4.3. Electrical Measurements

To characterize the devices, four-probe measurements are conducted. The write/read current pulses are applied through current channel and voltage difference is measured between voltage terminals. Current pulses are generated by Keithley 6221. The read current is set to be 50µA so as not to disturb the device state. The voltage difference is due to anomalous Hall effect (AHE) and measured by Keithley 2000 Multimeter. Hall resistance is defined by,

$$R_{AHE} = \frac{V_{AHE}}{I_{Channel}} \qquad (6)$$

Here, $V_{AHE}$ is the voltage difference across the two voltage terminals, and $I_{Channel}$ is the read current pulse amplitude flowing through the current channel.

### 4.4. Magnetic Anisotropy Field Estimation

To estimate the magnetic anisotropy field, in-plane magnetic field sweeping measurements can be conducted. The device magnetization gently tilts under the applied in-plane field, which causes a small deviation in Hall resistance. The deviating Hall resistance results in a bending hysteresis loop, which can be used to estimate magnetic anisotropy field. For details, refer to the Supporting Information Text and Supporting Figure S1.

### 4.5. Simulation methodology for co-design approach evaluation

The network used for the hardware-software co-design analysis was adopted from [59]. The network was trained for 40 epochs on the MNIST handwritten digit recognition dataset, with batch size of 100, a learning rate of 2.0 and momentum and dropout fractions of 0.5. After training, the network's neurons were converted to their spiking counterpart, i.e., to the spin stochastic neurons for inference. The input images were presented to the network as Poisson spike trains of length 50. A maximum spiking rate of 1000 was chosen. For the variational analysis, the bias current and the weights were randomly varied within the variation limit.

**4.6. Hardware-in-loop training methodology**

For the hardware-in-loop training, the images are converted to Poisson spike trains of length 100 time-steps. The feedforward network structure consists of 784×4 weight connections, which are randomly initialized. After modulation by the network, the impulses are scaled and biased according to the device characteristics obtained during characterization. A LabVIEW interface handles the generation of the proper pulsing scheme for the neuron devices with the help of the pulse current generator, Keithley 6221. The resultant Hall resistance is measured using a multimeter, Keithley 2000. Based on the resistance values after the reset and programming pulses, the interface counts the number of switching events for a single image. This is used to calculate the activation, by dividing it with the total time-steps, 100. The activation is then used for error and gradient calculations, followed by backpropagation through the network to ultimately calculate the weight updates for the feedforward network in software. The step is repeated for all the training images. During inference, the feedforward network is used, and the activations indicate the network's ability to accurately identify the patterns. The confidence is calculated from the normalized inputs to the neurons from the network and the neuron with the highest confidence is assigned to that class.

**Figures**

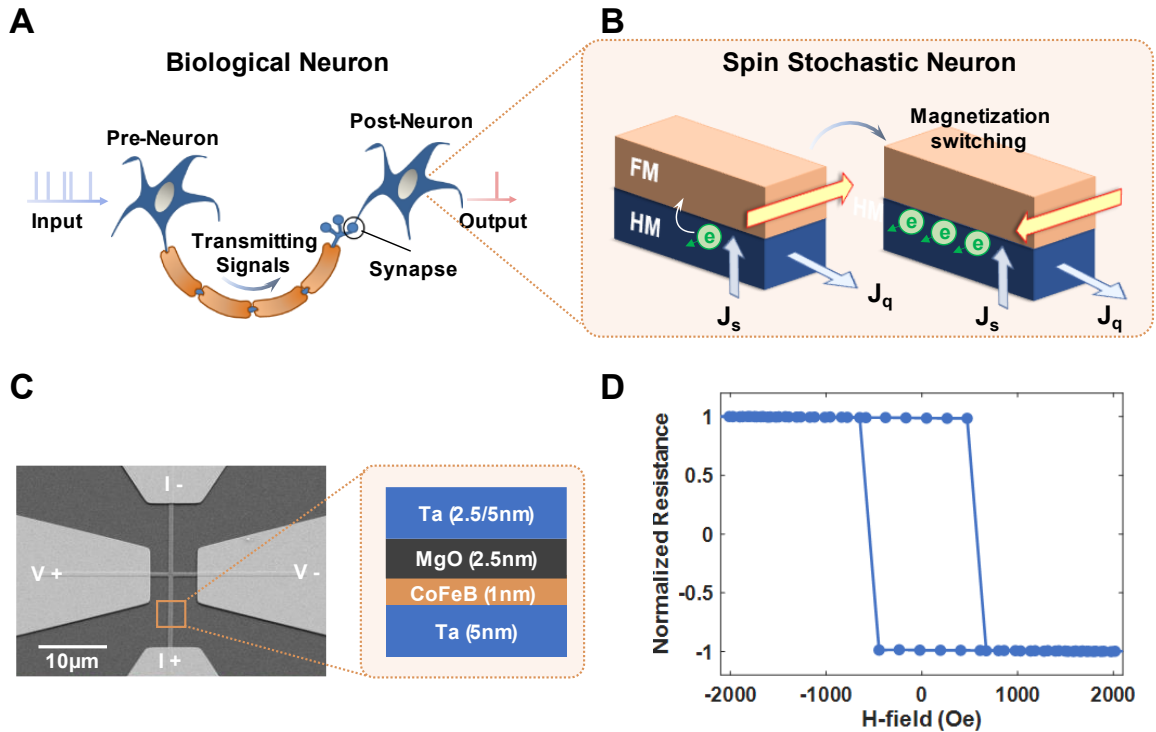

**Figure 1. Spin-orbit torque-based stochastic neuron.** (a) Schematic of biological neurons interconnected via synapses. Incoming input, propagated from pre-synaptic neurons and modulated by synapses, are integrated in the post-synaptic neuron, before generating an output spike. (b) Proposed spin stochastic neuron operation based on spin-orbit torque (SOT). Current through the heavy metal (HM) layer induces spin current, which exerts SOT on the ferromagnetic (FM) layer's magnetization. Switching of the device occurs when the charge current reaches the switching current, similar to a biological neuron. Thermal noise makes the switching stochastic. (c) SEM image of the device structure. Connections for measurement are annotated. The figure inset shows the material stack used for the device. (d) Magnetic hysteresis loop of a representative device with out-of-plane magnetic field (*H* field, *Oe*). The rectangular shape of the loop indicated perpendicular magnetic anisotropy (PMA) of the devices.

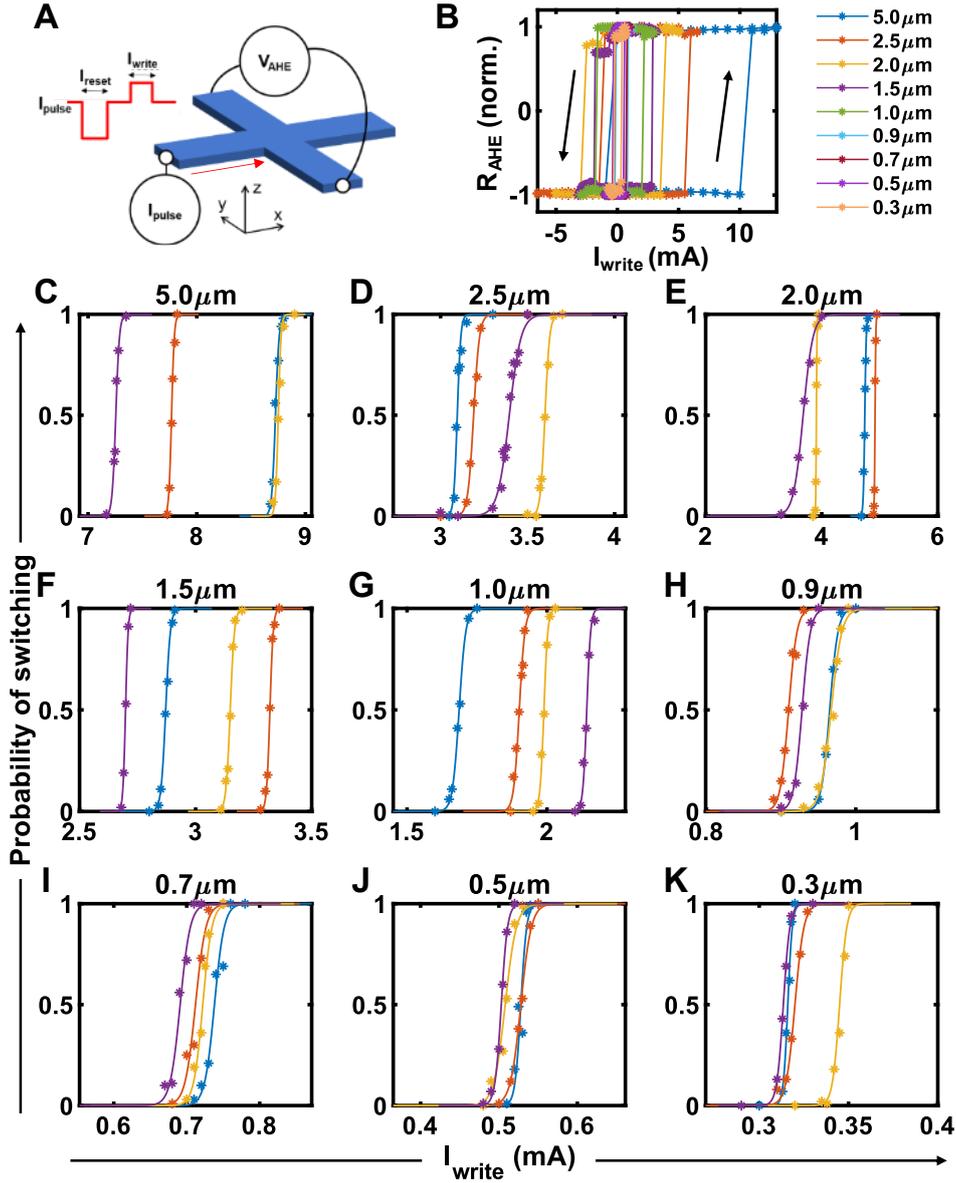

**Figure 2. Neuronal dynamics characterization.** (a) Four-probe measurement setup for characterizing. Programming (reset, write) and read pulses are applied along x-axis, while the anomalous Hall voltage is measured along the y-axis. An in-plane magnetic field is also applied. Inset shows the current pulse applied. In each iteration, a 100μs reset pulse and a 100μs write pulse are applied. Read pulses of 500ms and 50μA are applied after each programming pulse to read the device state. The interval between pulses is 2s. (b) Normalized Hall resistance of the various sized devices as the write current is gradually swept. We find that for sufficiently high switching current, the device switches from "-1" state to the "+1" state abruptly. We found that the hysteresis loop became larger with increasing device size. (c-k) The experimental results of 4 devices each of different sizes of spin neuron devices (5μm, 2.5μm, 2μm, 1.5μm, 1μm, 0.9μm, 0.7μm, 0.5μm, 0.3μm). Each device's switching dynamics is fitted to that of a sigmoid, showing close resemblance.

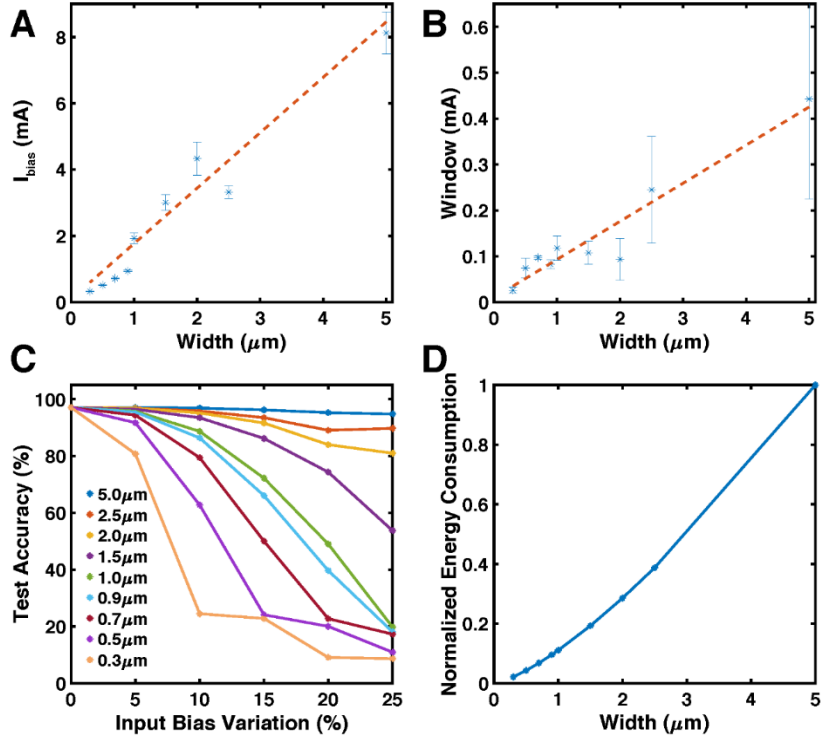

**Figure 3. Impact of device dimension on neural dynamics.** (a) Relationship between the bias current, $I_{bias}$ and device width. Switching current increases linearly with hall bar width. (b) Relationship between programming window and bar width. Again, with decreasing bar size, we observed that the programming window also decreases. Additionally, we observe that for the different sizes of the hall bars, we can have up to 25% variation from one device to another. (c) Effect of input bias current variation on network performance. It can be seen that the larger devices are able to handle the variations better in comparison to the scaled devices. (d) Relationship between normalized synaptic read energy required to achieve the target accuracy and device width. The energy consumption drops significantly as scaled devices allow for lower operating voltages for the synaptic crossbar. The system level energy is calculated as the sum of the read energy consumption of the synaptic arrays of all the layers in the network.

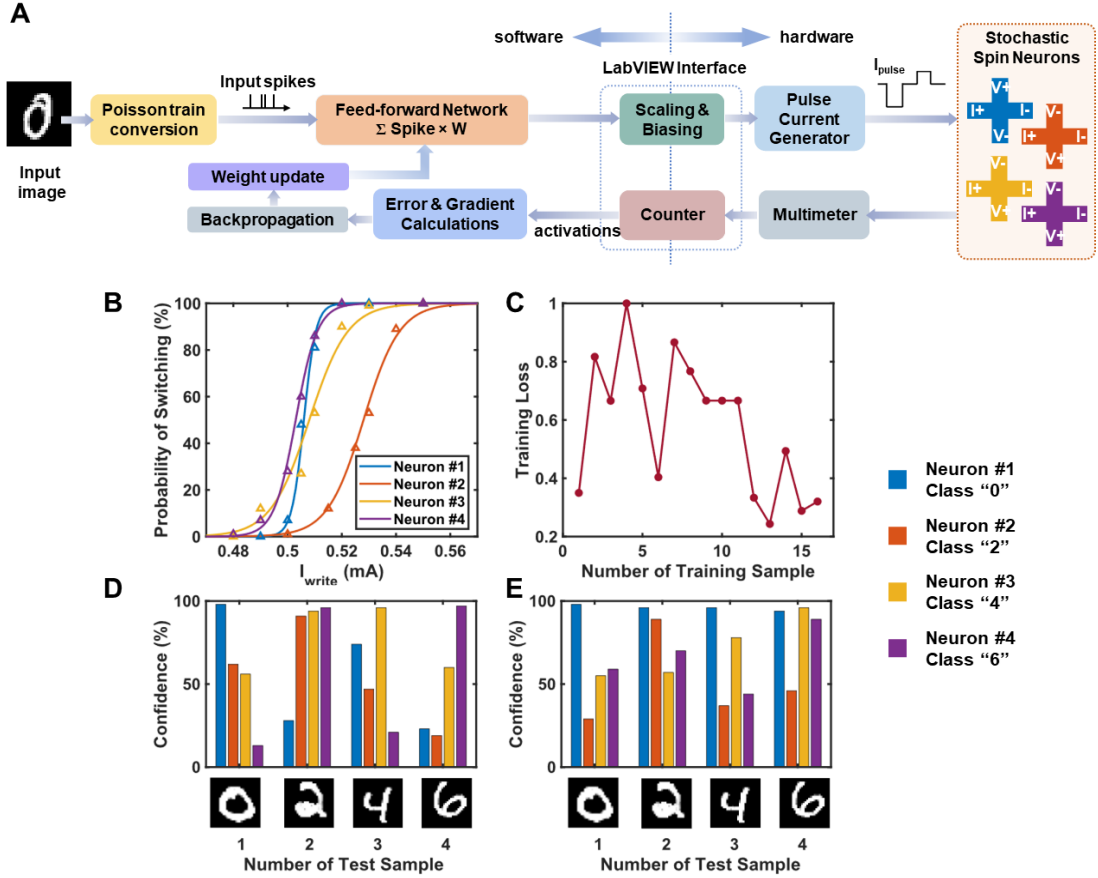

**Figure 4. Hardware-in-loop learning for the SOT Stochastic Spin Neuron Devices.** (a) The schematic of the hardware-in-loop (HIL) learning setup used for the experiments. See SI for detailed explanation. (b) Neuronal dynamics of the four-spin stochastic neurons used for HIL training. As can be seen, the four neurons have differing biases and operation windows. Note, each neuron represents a different class and is represented with a distinct color. (c) The training loss of the network over the 16 training images from (4 each from four classes, '0', '2', '4', '6') the MNIST Handwritten Digit Dataset. The training loss becomes low indicating the network is gradually learning. (d) The testing results on 4 test images from the MNIST dataset when HIL is used. The network is able to successfully classify 3 out of the 4 digits. The confidence is calculated from the normalized inputs to the neurons from the network and the neuron with the highest confidence is assigned that corresponding class. (e) The testing results on 4 test images from the MNIST dataset when HIL is not used. The software-trained network is only able to classify one image correctly, highlighting the need for including the hardware in training.

**Supporting Information**

Supporting Information is attached at the end of the manuscript.

**Acknowledgements**

We are thankful to N. Samarth, Y. Ou, and W. Yanez from the Department of Physics, Penn State University for their valuable suggestions regarding the fabrication of spin neuron devices.

**Author Contributions**

A.N.M.N.I., K.Y., and A.S. conceived the study. P.K., B.W.Z and W.G.W fabricated the thin films. A.N.M.N.I., K.Y., and A.K.S. fabricated the devices. A.N.M.N.I. and K.Y. measured the device characteristics and analyzed the results. A.N.M.N.I. and K.Y. carried out the hardware-in-loop experiments. A.N.M.N.I simulated the network-level effects and analyzed the results. A.N.M.N.I and A.S. prepared the manuscript. All authors participated in discussing the results and providing sections and comments for the manuscript.

A.N.M.N.I. and K.Y. contributed equally to this work.

**Funding**

This work was supported in part by the National Science Foundation under Grant ECCS 2028213 (fabrication and characterization of spin devices, implementation of hardware-in-loop training), Grant CCF 1955815 (spiking neural network computing framework) and DMR 1905783 (voltage controlled antiferromagnetism in magnetic tunnel junctions).

**Conflict of Interest Disclosure**

All authors declare that they have no competing interests.

**Data availability statement**

All data needed for evaluation are available in the main manuscript text and the supporting information provided.

K. Warrilow, J. P. Wang, W. G. Wang, *Sci. Rep.* **2023**, *13*, 1.

[59] R. B. Palm, *Tech. Univ. Denmark* **2012**, *5*.

# Supporting Information

**Hardware in Loop Learning with Spin Stochastic Neurons**

A N M Nafiul Islam, Kezhou Yang, Amit Kumar Shukla, Pravin Khanal, Bowei Zhou, Wei-Gang Wang, Abhronil Sengupta*

**Supporting Information Text**

**Magnetic Anisotropy Field Estimation**

To estimate the magnetic anisotropy field, in-plane magnetic field sweeping measurements can be conducted. The device magnetization gently tilts under the applied in-plane field, which causes a small deviation in Hall resistance, as is shown in Figure S1(a). The deviating Hall resistance results in a bending hysteresis loop shown in Figure S1(b), which can be used to estimate magnetic anisotropy field. The magnetization direction under in-plane field can be calculated by,

$$\frac{\partial E}{\partial \theta} = 0 \quad (S1)$$

Here, $E = -K_{eff} \cos^2 \theta - M_s H_x \cos\theta$ is the magnetic energy density. $H_x$ is the applied in-plane field, $\theta$ is the angle between magnetization direction $\vec{M}$ and $z$ direction, $K_{eff}$ is the perpendicular magnetic anisotropy (PMA) energy density and $M_s$ is the saturation magnetization. This leads to the following equation:

$$\frac{R_{AHE}}{R_0} = 1 - \frac{1}{2}\left(\frac{1}{H_{an}}\right)^2 H_x^2 \quad (S2)$$

Here, $H_{an} = \frac{2K_{eff}}{M_s}$ is defined as the anisotropy field, $R_0$ is the Hall resistance when $H_x = 0$ and $R$ is the Hall resistance during field sweeping. $H_{an}$ can be estimated by fitting the equation to the obtained data, as is shown in Figure S1(c). We find that the magnetic anisotropy field does not possess an obvious relation with bar size, as is shown in Figure S1(d).

**Micromagnetic Simulation**

To substantiate the linear relationship between device width and bias current and programming window, we conduct micromagnetic simulations. For this analysis, we assume mono-domain approximation. The magnetization dynamics of a mono-domain device switched by spin current can be described by the Landau-Lifshitz-Gilbert (LLG) equation:

$$\frac{d\hat{m}}{dt} = -\gamma(\hat{m} \times H) + \alpha\left(\hat{m} \times \frac{d\hat{m}}{dt}\right) + \frac{\mu_B}{qM_SV}(\hat{m} \times I_s \times \hat{m}) \quad (S3)$$

In the equation, $q$ is the charge of an electron, $M_S$ is the saturation magnetization, $V$ is the volume of the device, $\mu_B$ is the Bohr magneton, $\hat{m}$ is the normalized magnetization of the device, $H$ is the applied magnetic field and $I_s$ is the injected spin current.

We solve the LLG equations for various widths while keeping the length and thickness the same, i.e., $V$ is changing across the various devices. Note, the simulation parameters are taken from [21], and thus, do not correspond directly to the material parameters of our exact spin neuron devices. Despite that, as both material stacks obey the same physical equations, we can extrapolate a similar conclusion for our devices. The results of the simulation are given in SI Figure S7 and confirm the linear relation between device width and bias current and programming window.

| A | B |
|---|---|
| 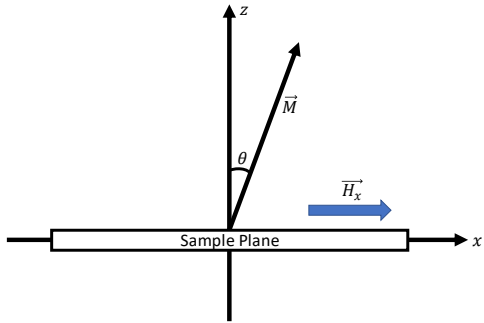 | 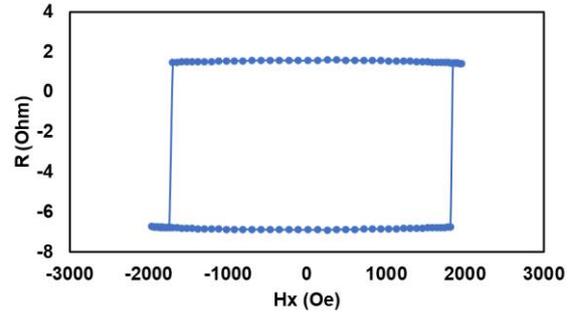 |

| C | D |
|---|---|
| 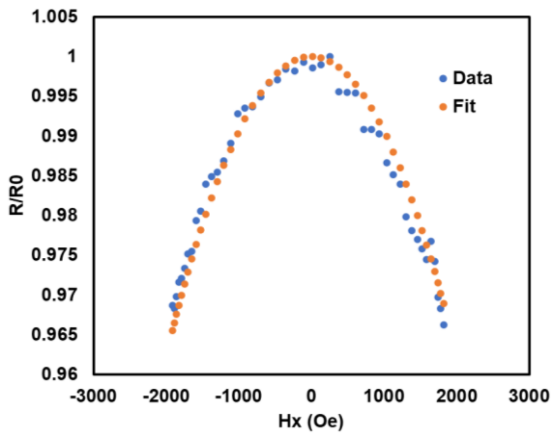 | 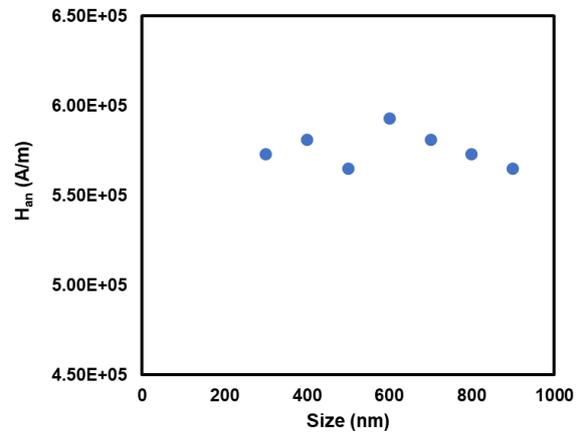 |

**Figure S1. Magnetic Anisotropy Field Estimation.** (**a**) Device magnetization under in-plane magnetic field. (**b**) Hysteresis loop of device under in-plane magnetic field sweeping (700nm Hall bar). The hysteresis loop bends due to the tilting of magnetization under in-plane magnetic field. The read current is 50μA. (**c**) Magnetic anisotropy field estimation by data fitting (700nm Hall bar). The fitted curve corresponds to $H_{an} = 5.8 \times 10^5 \text{A/m}$. (**d**) Magnetic anisotropy field of bars with different sizes.

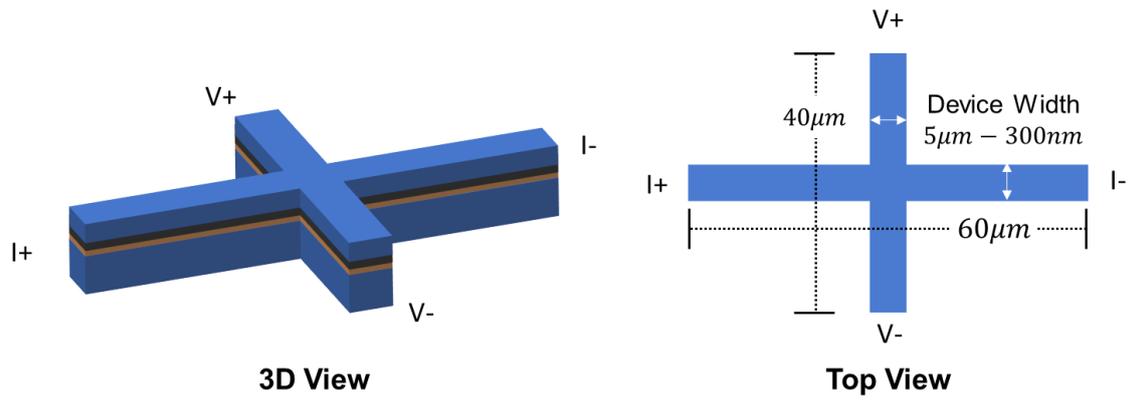

**Figure S2. Device Geometry.** The three-dimensional view is on the left and the top view is on the right. The dimensions are annotated in the top view. The width of both arms of the Hall bar are equal and represent the device bar width. The lengths of the Hall bar arms are chosen to ensure good contact with the metal contact pads.

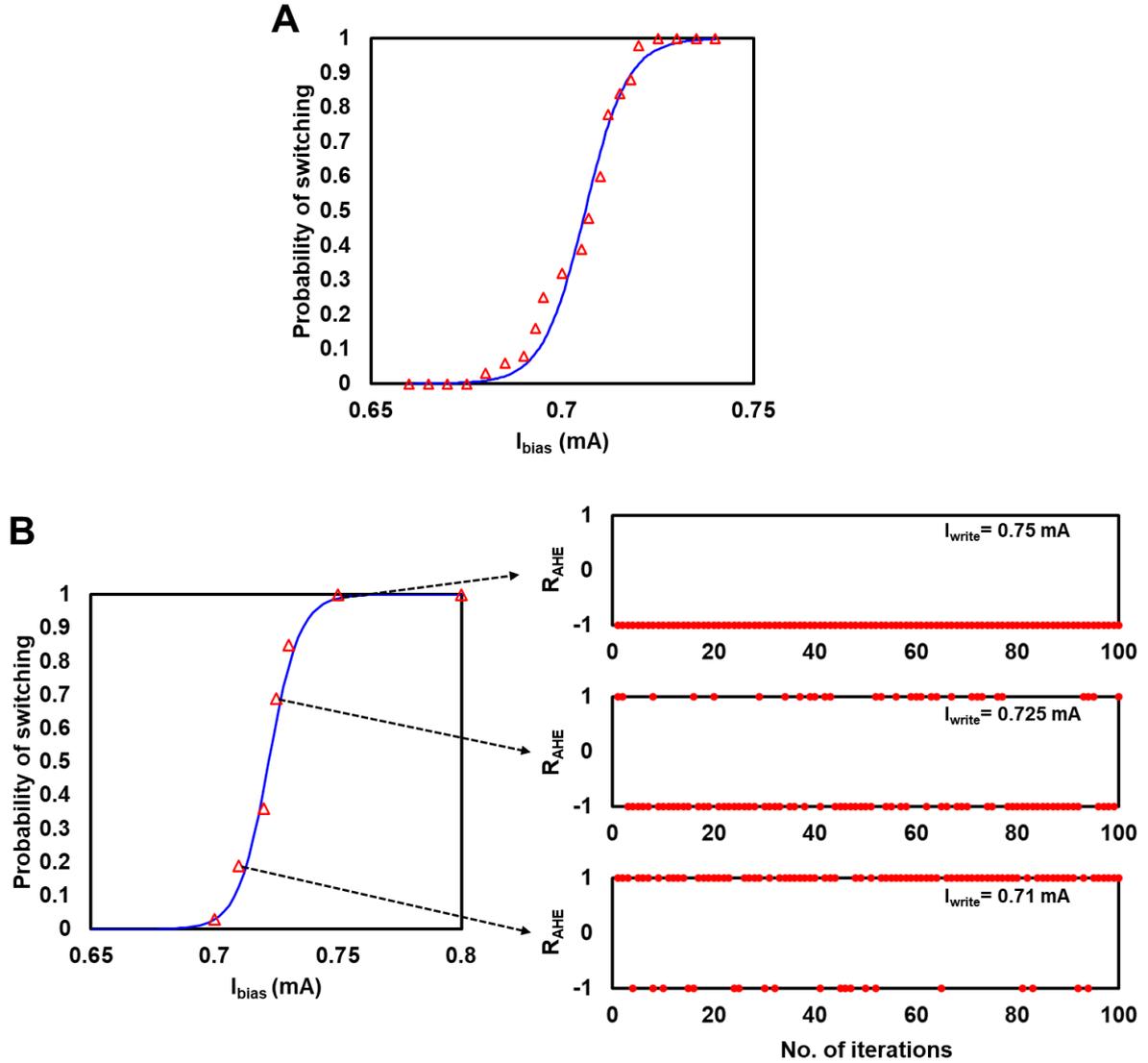

**Figure S3. Obtaining the sigmoidal characteristics of the neuronal devices. (a)** We confirm the sigmoidal nature of the switching probability of the spin neuron device by collecting switching data with small variations in the bias current. **(b)** We obtain each data point of the switching characteristics for each device by applying a specific write current for 100 iterations. Before each write current pulse, a reset current pulse is applied to reset the device to the high resistance state, as indicated by '+1'. After the write pulse is given, the state of the device is read again. If the device stays in the high resistive state ('+1'), there is no switching. If it goes to the low resistive state (indicated by '-1'), then the device is considered to be switched. We count the total number of switches in the 100 iterations to calculate the probability of switching of the devices.

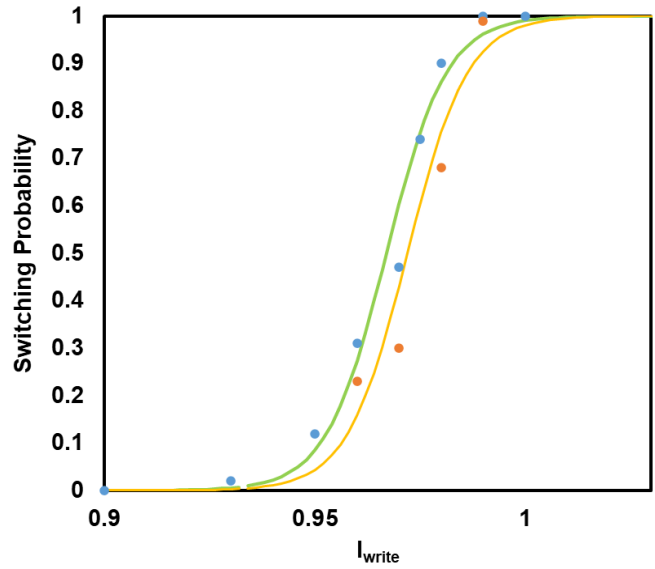

**Figure S4. Persistence of neuronal dynamics.** The neuronal dynamics of the same device was measured after a week, and it showed similar switching characteristics, with no significant variation (~0.5%) in the bias switching current.

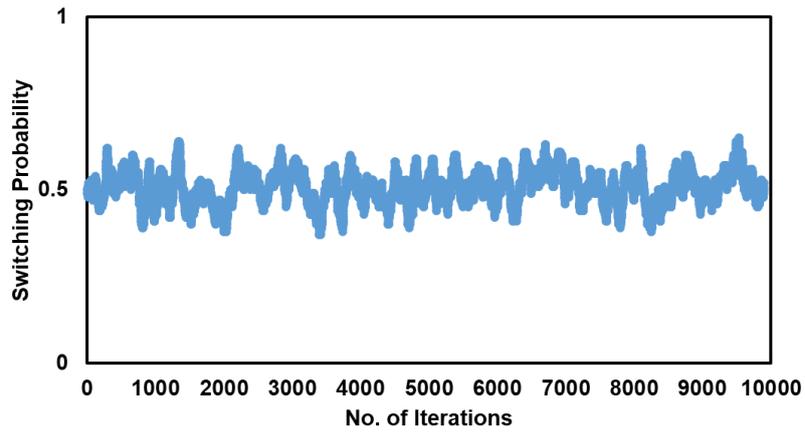

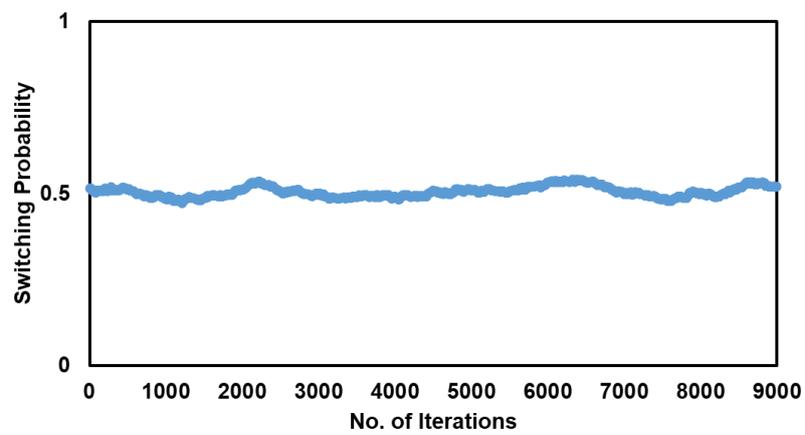

**Figure S5. Endurance test of neuronal dynamics.** We biased a spin neuron device at ~50% switching probability and repeated a reset and programming pulse cycle for 10,000 iterations. By taking the moving average over 100/1000 iterations, we can calculate the switching probability of the device. The device shows great consistency over the entire run, switching a total of 5061 times (equating to a 50.61% switching probability).

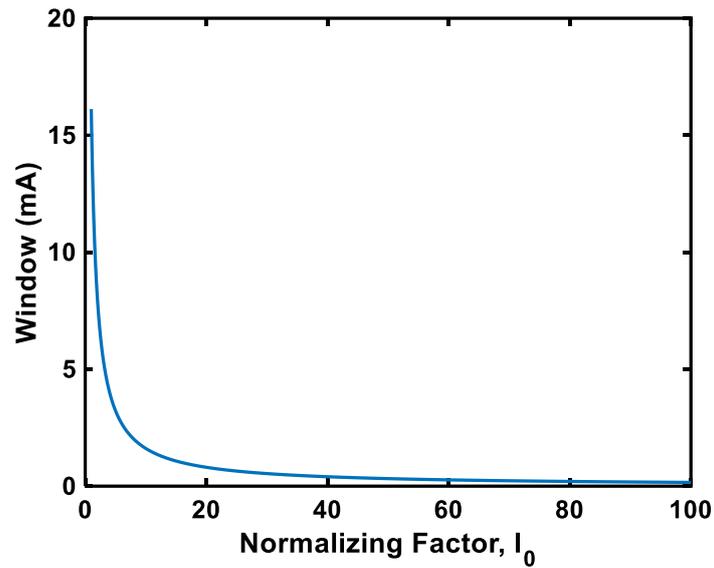

**Figure S6. Relationship between the dispersion metrics, $I_0$ and window.** The window is defined by the current pulse's amplitude range between which the switching probability goes from 0.01% and 99.9%. The factor $I_0$ is an unitless absolute number that normalizes the synaptic current such that the switching probability of the spin neuron devices resemble that of a sigmoid. As the device bar widths are reduced, the switching dynamics becomes less dispersed, i.e., the operating window is reduced and thus similarly the required normalizing factor is increased.

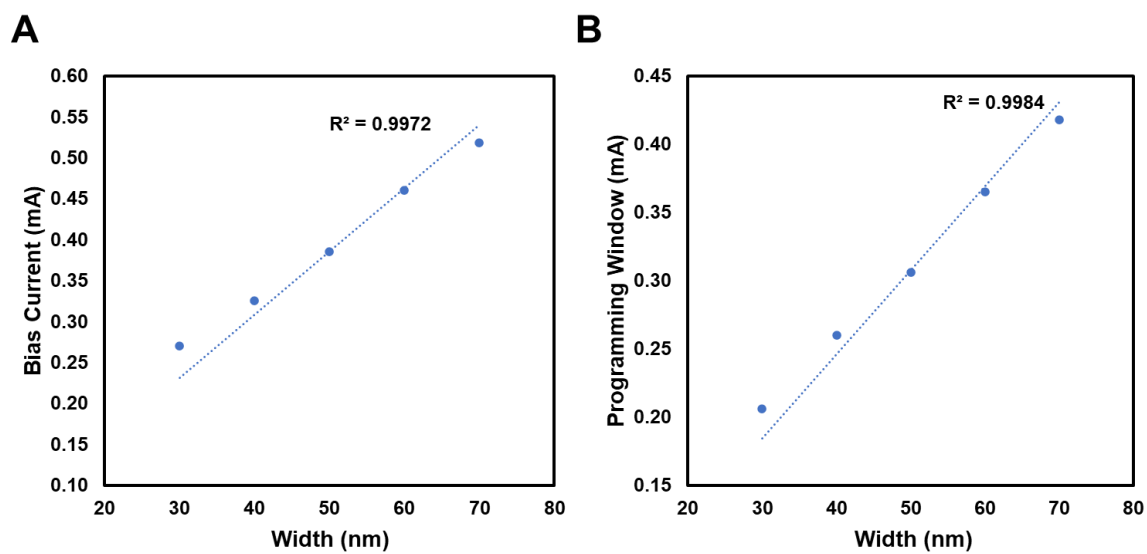

**Figure S7. Results of Micromagnetic Simulations.** The results confirm the linear relationship between device width and bias current **(A)** and device width and programming window **(B)**.

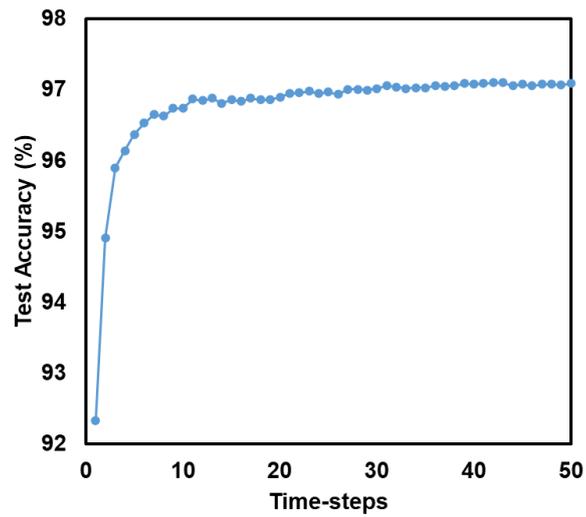

**Figure S8. Trade-off between latency and accuracy.** The test accuracy during inference with the input coming in as Poisson spike trains. As the number of time-steps is increased, the performance of the spiking neural network closely approximates that of the traditionally trained analog non-spiking neural network.

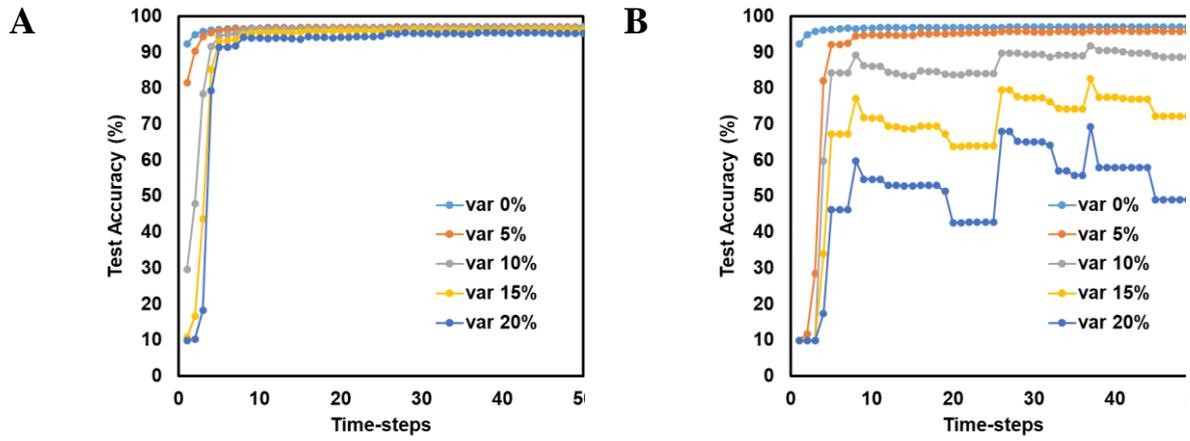

**Figure S9. Trade-off between latency and bias variation.** As the percentage of bias variation is increased in the network, the number of time-steps it takes to converge to the final accuracy increases. The case for 5µm **(A)** and 1µm **(B)** width devices are shown. While for the 5µm case, the network is able to converge closer to its original accuracy (accuracy without variation), for the 1µm case, the network behaves erratically, and performance becomes poorer as the variation is increased.

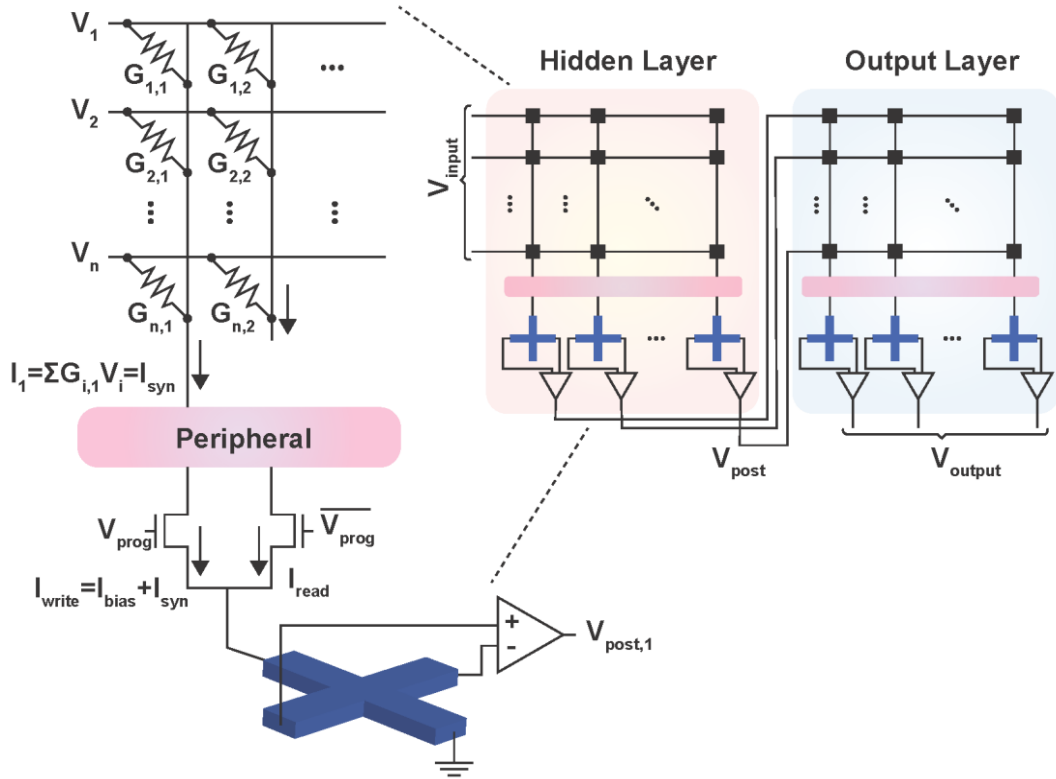

**Figure S10. System-scale implementation with the proposed neurons interfaced with cross-arrays of synaptic devices.** Each layer of the network can be embedded in crossbar arrays of resistive synaptic devices (can be potentially implemented by spin devices). If the layer size is large, it can be divided across multiple small crossbar arrays. The outputs of the hidden layers are fed to the next layer until the output layer. One column of the crossbar array is enlarged on the left to show its operation. Input voltages ($V_i$) get modulated by the conductances ($G_{i,j}$) and are summed according to Kirchhoff's law to generate the synaptic current. The peripheral circuitry adds the necessary bias current to create $I_{write}$ which programs the spin neuron, when $V_{prog}$ is high. A read current is applied to get the state of the device, when $V_{prog}$ is low. The output across the hall bar is fed to the next layer through a driving op-amp.

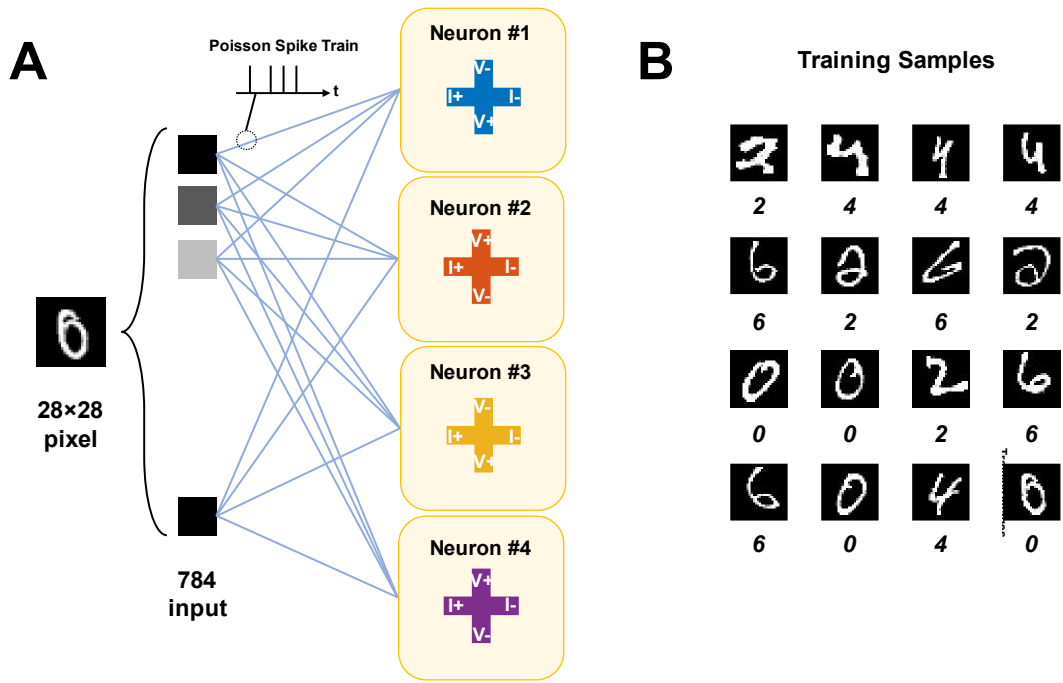

**Figure S11. Training for the hardware-in-loop scheme.** **(A)** The network architecture used for training in the hardware-in-loop scheme. **(B)** Training images used for hardware-in-loop learning. We used 4 classes ("0", "2", "4" and "6") with 4 images of each class.

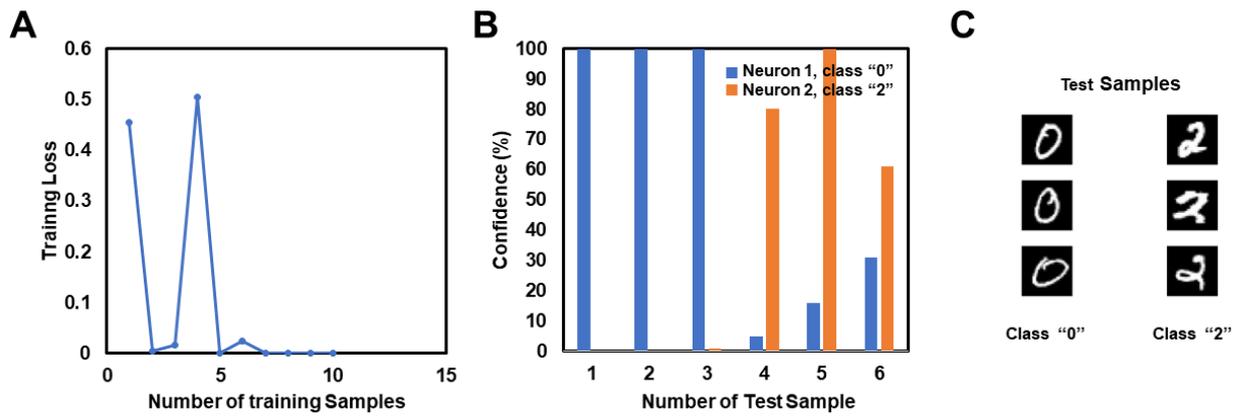

**Figure S12. Training a single device with time-multiplexing to emulate 2 neurons through hardware-in-loop scheme.** (**A**) The training loss of the network over the training images. (**B**) The testing results on 6 test images from the MNIST dataset. The network achieves an accuracy of 100%. (**C**) Test samples used for the inference of the two classes ("0" and "2") used.